\documentclass[runningheads]{llncs}
\usepackage[T1]{fontenc}
\usepackage{graphicx}
\usepackage{booktabs}
\usepackage{amsmath}
\usepackage{amsfonts}
\usepackage{algorithm}      
\usepackage{algpseudocode}  
\usepackage{appendix}

\usepackage{array}  
\usepackage{longtable}  

\begin{document}
\title{We Care Each Pixel: Calibrating on Medical Segmentation Model}
%
%
\author{Wenhao Liang,Wei Zhang, Lin Yue, Miao Xu, \\Olaf Maennel, Weitong Chen}

\institute{Adelaide University}
\maketitle              
\begin{abstract}
Medical image segmentation is fundamental for computer-aided diagnostics, providing accurate delineation of anatomical structures and pathological regions. While common metrics such as Accuracy, DSC, IoU, and HD primarily quantify spatial agreement between predictions and ground-truth labels, they do not assess the calibration quality of segmentation models, which is crucial for clinical reliability. To address this limitation, we propose \emph{pixel-wise} Expected Calibration Error (pECE), a novel metric that explicitly measures miscalibration at the pixel level, thereby ensuring both spatial precision and confidence reliability. We further introduce a morphological adaptation strategy that applies morphological operations to ground-truth masks before computing calibration losses, particularly benefiting margin-based losses such as Margin SVLS and NACL. Additionally, we present the \emph{Signed Distance Calibration Loss} (SDC), which aligns boundary geometry with calibration objectives by penalizing discrepancies between predicted and ground-truth signed distance functions (SDFs). Extensive experiments demonstrate that our method not only enhances segmentation performance but also improves calibration quality, yielding more trustworthy confidence estimates. Code is available at: \url{https://github.com/EagleAdelaide/SDC-Loss}

\end{abstract}

\section{Introduction}
Medical image segmentation entails more than simply delineating boundaries~\cite{wang2022boundary,rogowska2009overview,lee2020structure,pham2000current}; it requires models to learn clinically meaningful representations that can inform diagnostic and prognostic decisions. In practice, segmentation models must provide both high spatial accuracy and reliable confidence estimates, because overconfident predictions may mislead clinical decision-making~\cite{sox2024medical}. Although overconfidence can reduce overlap errors in certain cases, it substantially increases calibration error if predicted probabilities fail to reflect true uncertainty~\cite{guo2017calibration,kumar2019verified}. In classification tasks, well-calibrated models ensure that predicted probabilities align with actual outcome frequencies. For instance, if a pixel is assigned a probability of 0.8 for belonging to a lesion, it should indeed be part of a lesion 80\% of the time over many samples~\cite{guo2017calibration}. However, even models with high Dice or IoU scores can suffer from poor calibration in segmentation tasks~\cite{yeung2022unified,yeung2023calibrating}, compromising their clinical utility. Accurate probability estimates are crucial in medical applications, influencing decisions about follow-up imaging, treatment planning, and risk assessment~\cite{hricak2007imaging}.

To measure calibration in segmentation, one must evaluate predictions at the pixel or voxel level, because each pixel represents an independent classification decision. While Expected Calibration Error (ECE) is widely used for classification, applying it to large-scale pixel-based tasks necessitates adaptation. Therefore, we propose \emph{pixel-wise ECE} (pECE), which extends calibration analysis to the considerable number of pixel-level predictions in medical images. Our method bridges the gap between high segmentation accuracy and reliable uncertainty estimation, thereby promoting safer automated analysis in clinical settings. Specifically, our contributions are threefold:
\begin{itemize}
    \item We present a novel \textbf{Signed Distance Calibration (SDC) Loss} (\S\ref{subsec:sdf_loss}), which integrates cross-entropy, localized calibration regularization, and signed distance function (SDF) regression into a unified objective that simultaneously addresses boundary accuracy and predictive confidence.
    \item We propose a \textbf{Spatially Adaptive Margin Module with Morphological Transforms} (\S\ref{subsec:adaptive_margin}), which augments local target distributions through morphological operations on ground-truth masks, enhancing boundary delineation robustness and mitigating label noise.
    \item We introduce a \textbf{pixel-wise Expected Calibration Error (pECE)} (\S\ref{subsec:pece}), designed for high-resolution calibration analysis and equipped with a built-in penalty for false positives in critical regions, thereby producing more reliable confidence estimates.
\end{itemize}

\section{Related Work}

Deep neural networks have made substantial progress in medical image segmentation~\cite{anwar2018medical,sarvamangala2022convolutional}, yet they often generate overconfident predictions that may compromise clinical utility~\cite{guo2017calibration}. To alleviate this problem, recent studies have emphasized model calibration to ensure that predicted confidence scores accurately represent true likelihoods. Early methods adopted label smoothing (LS)~\cite{szegedy2016rethinking} and confidence penalty techniques (ECP)~\cite{pereyra2017regularizing} to mitigate excessive certainty, showing improvements in both calibration and generalization. Meanwhile, Focal Loss (FL)~\cite{lin2017focal} was developed to address class imbalance by emphasizing more challenging samples; however, its effect on calibration remained limited without further adjustments~\cite{mukhoti2020calibrating}. Subsequent approaches have investigated spatially adaptive label smoothing, such as Spatially Varying Label Smoothing (SVLS)~\cite{islam2021spatially}, and margin-based regularizers like Margin-based Label Smoothing (MbLS)~\cite{liu2022devil}, both aiming to refine predicted probability distributions and reduce miscalibration. In parallel, methods like Neighbor Aware Calibration Loss (NACL)~\cite{murugesan2024neighbor} and Focal Calibration Loss (FCL)~\cite{liang2024calibrating} leverage local structural information to adapt to varying confidence levels across different regions. Nevertheless, a common limitation among these approaches is the reliance on uniform penalty weights, which overlook class-specific or region-specific uncertainties, potentially diminishing calibration quality in complex scenarios. Moreover, most methods still evaluate calibration errors at a global level, risking the omission of subtle pixel-level inconsistencies that can be critical for clinical decisions.

\begin{figure}
\includegraphics[width=\textwidth]{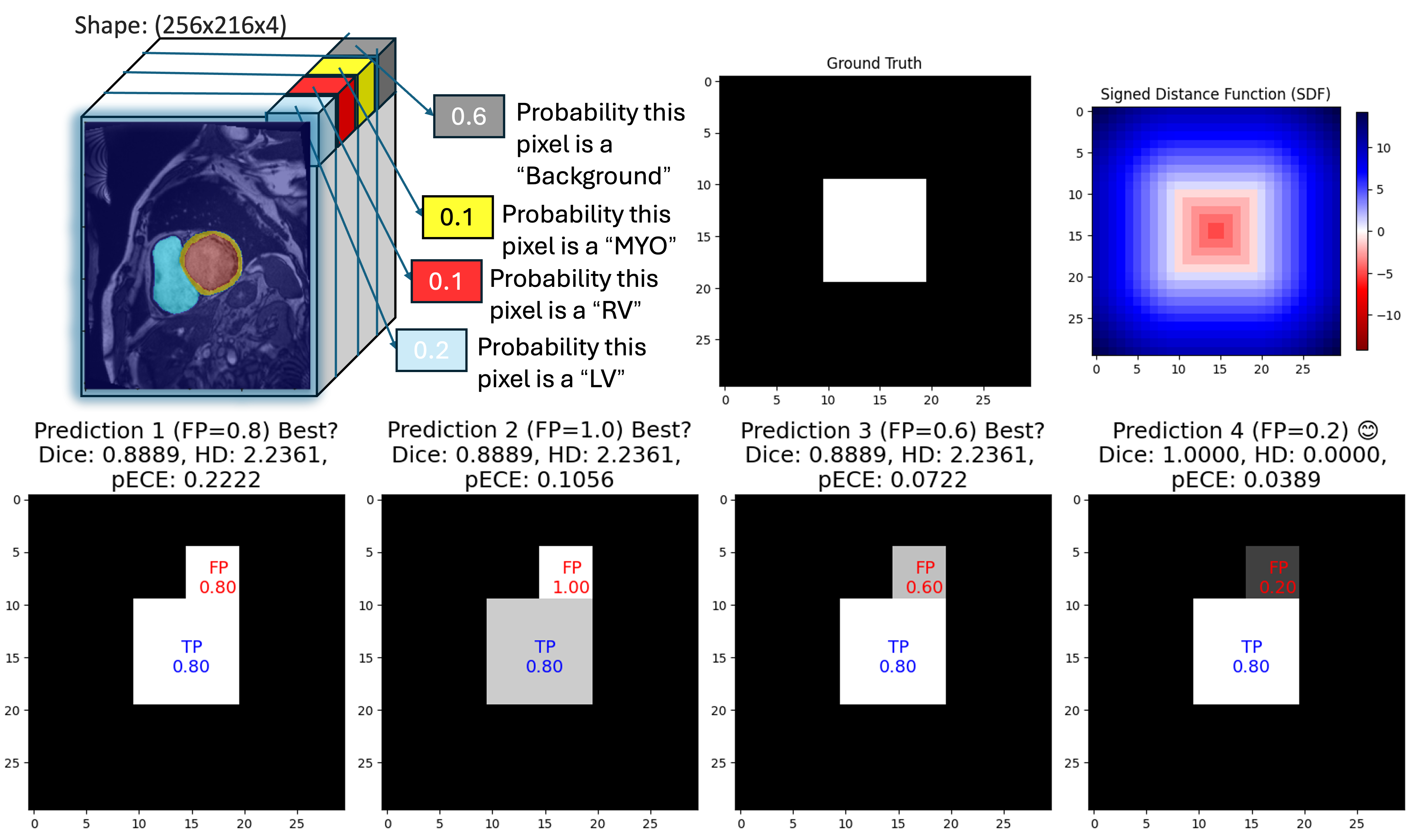}
\caption{Evaluation of segmentation predictions on the ACDC dataset, comparing Dice Score (Dice), Hausdorff Distance (HD), and pixel-wise Expected Calibration Error (pECE). The top row illustrates the dataset structure, pseudo ground-truth masks, and the signed distance function (SDF). The bottom row presents four pseudo segmentation predictions, highlighting True Positives (TP) and False Positives (FP) while illustrating variations in segmentation accuracy and calibration error. Prediction 4 achieves the lowest pECE, indicating optimal segmentation confidence.}
\label{pECE}
\end{figure}

\section{Methodology}

\begin{figure}[ht]
\includegraphics[width=\textwidth]{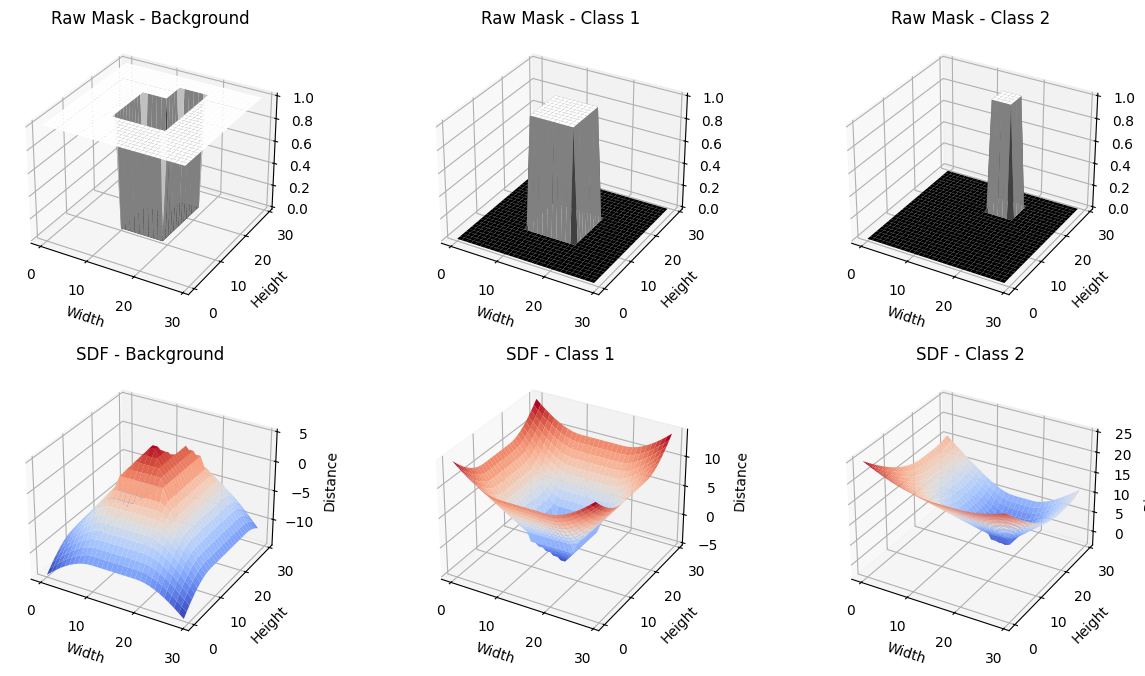}
\caption{Illustration of raw binary segmentation masks (top row) and their corresponding Signed Distance Function (SDF) representations (bottom row). Negative distances (inside the object) are shown in blue, while positive distances (outside the object) are shown in red. SDFs provide a continuous representation of object boundaries, enhancing spatial awareness for segmentation models.}
\label{mask_sdf}
\end{figure}

\subsection{Signed Distance Calibration (SDC) Loss}
\label{subsec:sdf_loss}
\begin{figure}[ht]
\centering
\includegraphics[width=\textwidth]{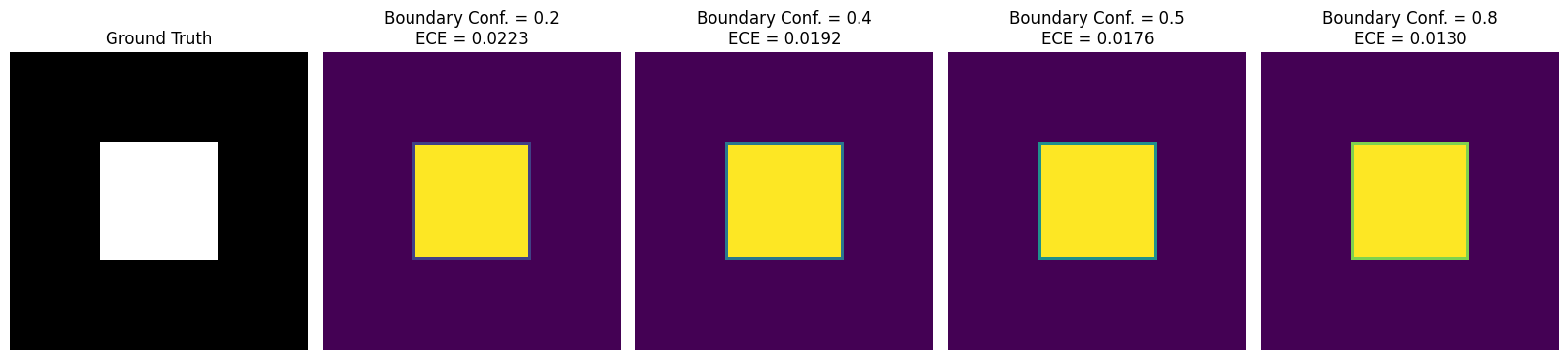}
\caption{Segmentation predictions with varying boundary confidence and their ECE.}
\label{conf_ece}
\end{figure}

In segmentation tasks, models typically exhibit high confidence in regions where objects are clearly present (or absent), while boundary regions often contain inherent ambiguity. However, many networks remain overly confident at these boundaries, creating a mismatch between predicted confidence and actual accuracy (Fig.~\ref{conf_ece}). This mismatch raises the Expected Calibration Error (ECE). Increasing boundary sensitivity encourages the model to moderate its confidence at boundaries, leading to probability estimates that more accurately capture true uncertainty and thus reduce overall calibration error.

To improve boundary sensitivity and calibration, we propose the \emph{Signed Distance Calibration} (SDC) loss, which combines three complementary components into a single objective function. First, the standard cross-entropy loss, \(\mathcal{L}_{\text{CE}}(\mathbf{z}, \mathbf{y})\), enforces pixel-level label fidelity. Here, \(\mathbf{z} \in \mathbb{R}^{B \times C \times H \times W}\) denotes the raw logits for a batch of \(B\) images, \(C\) is the number of classes, and \(\mathbf{y} \in \mathbb{R}^{B \times H \times W}\) is the ground-truth mask. The predicted probabilities are given by \(\mathbf{p} = \text{softmax}(\mathbf{z})\). Next, a local calibration term \(\mathcal{L}_{\text{conf}}(\mathbf{p}, \tilde{\mathbf{y}})\) measures the discrepancy—using an \(\ell_1\) or \(\ell_2\) norm—between \(\mathbf{p}\) and a locally smoothed target \(\tilde{\mathbf{y}}\) obtained via neighborhood-based filtering (e.g., mean or Gaussian). This term aligns model predictions with local ground-truth structure and promotes spatially coherent confidence estimates. Lastly, a signed distance function (SDF) penalty \(\mathcal{L}_{\text{SDF}}(\hat{\mathbf{s}}, \mathbf{s})\) enforces boundary precision. Let \(\mathbf{s}\) denotes the SDF derived from the ground-truth mask \(\mathbf{y}\), and \(\hat{\mathbf{s}}\) is the corresponding prediction. Unlike traditional losses that may overlook geometric details, this penalty emphasizes discrepancies near object boundaries. Altogether, the SDC loss is defined as

\begin{equation}
\label{eq:sdc_loss}
\begin{aligned}
\mathcal{L}_{\mathrm{SDC}}
&=
\underbrace{\mathcal{L}_{\mathrm{CE}}(\mathbf{z}, \mathbf{y})}%
_{\substack{\text{Cross-Entropy} \\ \text{(pixel fidelity)}}}
\;+\;
\alpha\,\underbrace{\mathcal{L}_{\mathrm{conf}}(\mathbf{p}, \tilde{\mathbf{y}})}%
_{\substack{\text{Local Calibration Term} \\ \text{(spatial coherence)}}}
\;+\;
\lambda_{\mathrm{SDF}}\,\underbrace{\mathcal{L}_{\mathrm{SDF}}(\hat{\mathbf{s}}, \mathbf{s})}%
_{\substack{\text{SDF Penalty} \\ \text{(boundary precision)}}}.
\end{aligned}
\end{equation}

where \(\alpha\) and \(\lambda_{\text{SDF}}\) control the relative influence of the local calibration term and the SDF penalty, respectively. The SDF component thus acts as a boundary-sensitive regularizer rather than a full mask reconstruction, guiding the model to learn both spatially coherent and geometrically informed confidence maps. 

\subsection{Spatially Adaptive Margin With Morphological Transforms}
\label{subsec:adaptive_margin}
Although conventional local smoothing helps mitigate overconfidence, it often ignores label noise and boundary inconsistencies. To address these issues, we apply a morphological operator \(\mathcal{M}\) to the ground-truth mask \(\mathbf{y}\), producing a refined mask \(\mathbf{y}_m = \mathcal{M}(\mathbf{y})\). Specifically, when \(\mathcal{M}\) is the \emph{morphological gradient} of a binarized mask \(A\), it is defined as
\(\label{eq:morph_gradient}
    \mathbf{y}_m
    \;=\;
    \mathrm{Grad}(A)
    \;=\;
    (A \oplus B)
    -
    (A \ominus B),
\)
where \(B\) is a structuring element (e.g., \(3 \times 3\)) that either suppresses noise (erosion) or sharpens boundaries (dilation). We then smooth \(\mathbf{y}_m\) with a neighborhood-based filter (e.g., mean or Gaussian) to obtain the locally aggregated target map \(\tilde{\mathbf{y}}_{m}\). Let \(\mathbf{p} = \text{softmax}(\mathbf{z}) \in \mathbb{R}^{B \times C \times H \times W}\) denote the predicted probabilities. The spatially adaptive margin loss is formulated as
\begin{equation}\label{eq:margin_svls}
    \mathcal{L}_{\text{margin}}(\mathbf{z}, \mathbf{y}_m)
    =
    \underbrace{\mathcal{L}_{\text{CE}}(\mathbf{z}, \mathbf{y}_m)}_{\text{cross-entropy}}
    +
    \alpha\,
    \underbrace{\bigl\|\mathbf{p} - \tilde{\mathbf{y}}_{m}\bigr\|}_{\text{margin term}},
\end{equation}
where \(\alpha\) balances the standard cross-entropy term against local calibration. By incorporating morphological operations into the local smoothing procedure, we selectively regularize boundary regions, mitigating label noise and prompting the network to learn boundary-aware probability maps. 

\subsection{Pixel-wise Expected Calibration Error (pECE)}
\label{subsec:pece}
Global calibration metrics, such as overall ECE, can conceal local miscalibration by aggregating errors across an entire image. We propose a \emph{pixel-wise} ECE (\textbf{pECE}) that computes calibration quality on a per-pixel basis. The range \([0,1]\) is divided into \(B\) bins, and each pixel is assigned to a bin according to its predicted confidence. Let \(\bar{p}_{b}\) and \(\bar{a}_{b}\) represent the average confidence and accuracy in bin \(b\), and let \(\text{FPConf}_{b}\) be the mean confidence of false-positive pixels in that bin. We define pECE as:
\begin{equation}\label{eq:pece}
    \text{pECE}
    =
    \sum_{b=1}^{B}
    \frac{
       \bigl|\,
         (\bar{p}_{b} - \bar{a}_{b})
         +
         w_{\text{fp}} \cdot \text{FPConf}_{b}
       \bigr|
       \,\times\,
       |\Omega_{b}|
    }{|\Omega|},
\end{equation}
where \(\Omega_{b}\) denotes the set of pixels in bin \(b\), \(|\Omega_{b}|\) its size, and \(|\Omega|\) the total number of pixels. The coefficient \(w_{\text{fp}}\) penalizes overconfident false positives, highlighting whether particular regions or boundaries are systematically miscalibrated. This localized approach is especially important in clinical applications, where a small but critical miscalibration (e.g., at a lesion boundary) can lead to significant diagnostic or treatment errors. 
\section{Experiments}

\noindent \textbf{Datasets.}
We evaluate our approach on four publicly available medical imaging datasets, largely following the protocol outlined by Neighbor Aware Calibration Loss (NACL)~\cite{murugesan2024neighbor}. \noindent \textbf{ACDC}~\cite{bernard2018deep}: This cardiac MR dataset comprises 100 volumes with pixel-wise annotations. As in previous work, we convert the volumes into 2D slices, resize them to \(224 \times 224\), and use these processed slices for both training and inference. \noindent \textbf{FLARE}~\cite{Ma2021postrank}: This abdominal CT dataset consists of 360 scans with multi-organ labels. Each volume is resampled to a uniform spatial resolution, then cropped to \(192 \times 192 \times 30\), ensuring consistency in both the training and testing phases. \noindent \textbf{BraTS 2019}~\cite{Menze2015TheBRATSJ,Bakas2017AdvancingFeaturesJ,Bakas2018IdentifyingChallengeJ}:  We further include the Brain Tumor Segmentation (BraTS) 2019 challenge dataset, featuring 335 multi-channel MRI scans (FLAIR, T1, T1-contrast, and T2) with glioma segmentation masks. \noindent \textbf{PROSTATE}~\cite{antonelli2022medical}:
Finally, we consider the PROSTATE subset from the Medical Segmentation Decathlon (MSD) containing 32 MRI volumes.

\noindent \textbf{Baselines.} 
We compare our method against several established calibration and state-of-the-art losses, including Focal Loss (FL)~\cite{lin2017focal}, Entropy-based Confidence Penalty (ECP)~\cite{pereyra2017regularizing}, Label Smoothing (LS)~\cite{szegedy2016rethinking}, Spatially Varying Label Smoothing (SVLS)~\cite{islam2021spatially}, Margin-based Label Smoothing (MbLS)~\cite{liu2022devil}, Neighbor Aware Calibration Loss (NACL)~\cite{murugesan2024neighbor}, and Focal Calibration Loss (FCL)~\cite{liang2024calibrating}. As segmentation backbones, we employ the widely used \textbf{U-Net}~\cite{ronneberger2015u} and \textbf{nnU-Net}~\cite{nnUNet} architectures. For fair comparisons, we adopt the hyperparameters from Tab.~\ref{tab:quantitative_performance}. Each model is trained for 100 epochs using the Adam optimizer~\cite{kingma2014adam} with a batch size of 16. The learning rate is initially set to \(10^{-3}\) for the first 50 epochs and reduced to \(10^{-4}\) thereafter. Following~\cite{murugesan2024neighbor}, we train our models on 2D slices but evaluate on reconstructed 3D volumes. The best-performing model is determined by the highest mean DSC score on the validation set. We conduct 5-fold cross-validation and report the mean metric values across all folds for each baseline.

\noindent \textbf{Evaluation.}
We report two widely adopted segmentation metrics in medical imaging: Dice Similarity Coefficient (DSC) and the 95\% Hausdorff Distance (HD). To measure calibration quality, we follow~\cite{murugesan2022calibrating,murugesan2023trust,murugesan2024neighbor} and compute the Expected Calibration Error (ECE)~\cite{naeini2015ece} for foreground classes, as recommended by~\cite{islam2021spatially}, along with the Class-wise Expected Calibration Error (CECE)~\cite{kumar2019verified} (using a threshold of \(10^{-3}\)). In addition, we introduce the \textbf{pixel-wise ECE (pECE)}, which captures fine-grained calibration discrepancies at the pixel level. To fairly compare performance, we use the Friedman ranking~\cite{friedman1937use}. 


\begin{table*}
\centering
\caption{\textbf{U-Net Segmentation and Calibration Results\protect\footnotemark[1].} 
Shown are discrimination metrics (DSC~$\uparrow$, HD~$\downarrow$) alongside calibration measures (ECE~$\downarrow$, CECE~$\downarrow$, pECE~$\downarrow$). Bold and underlined values highlight the best and second-best scores, respectively.}

\label{tab:quantitative_performance}
\resizebox{\textwidth}{!}{%
\begin{tabular}{l|ccccc|ccccc|cc}
\hline
\toprule
& \multicolumn{5}{c|}{\textbf{ACDC}} & \multicolumn{5}{c|}{\textbf{FLARE}} & {\textbf{Friedman}} \\
\hline 
\textbf{} & \textbf{DSC} & \textbf{HD} & \textbf{ECE} & \textbf{CECE} & \textbf{pECE} & \textbf{DSC} & \textbf{HD} & \textbf{ECE} & \textbf{CECE} & \textbf{pECE} & \textbf{Rank} \\
\hline
\midrule
DiceCE \cite{milletari2016v}  & 0.828 & 3.14 & 0.137 & 0.084 & 0.457 & 0.716 & 9.7 & 0.076 & 0.049 & 0.774 & 7.70 (9)  \\
FL \cite{lin2017focal} ($\gamma = 3$)  & 0.620 & 7.30 & 0.153 & 0.179 & 0.224 & 0.834 & 6.65 & 0.053 & 0.059 & \underline{0.217} & 7.70 (8) \\
ECP \cite{pereyra2017regularizing} ($\lambda = 0.1$)  & 0.782 & 4.44 & 0.130 & 0.094 & 0.193 & 0.857 & 5.30 & 0.037 & \textbf{0.027} & 0.307 & 5.15 (5) \\
LS \cite{szegedy2016rethinking} ($\alpha = 0.1$)  & 0.809 & 3.30 & 0.083 & 0.093 & \underline{0.177} & 0.856 & 5.33 & 0.055 & 0.049 & \textbf{0.216} & 5.40 (6) \\
$SVLS_{IPMI'21}$ \cite{islam2021spatially} ($\sigma = 2.0$) & 0.824 & 2.81 & 0.091 & 0.083 & 0.179 & 0.857 & 5.72 & 0.039 & 0.036 & 0.420 & 4.60 (4)\\
$MbLS_{CVPR'22}$ \cite{liu2022devil} ($m = 10$) & 0.827 & 2.99 & 0.103 & 0.081 & 0.206 & 0.855 & 5.75 & 0.046 & 0.041 & 0.580 & 5.70 (7)\\
$NACL_{MICCAI'23}$ \cite{murugesan2023trust} & 0.854 & 2.93 & 0.068 & 0.061 & 0.183 & \textbf{0.859 }& 4.88 & \underline{0.031} & \underline{0.031} & 0.455 & 3.20 (2)\\
FCL \cite{liang2024calibrating} ($\gamma = 3$, $\lambda = 0.1$) & \underline{0.864} & \textbf{1.77} & \underline{0.052} & \underline{0.045} & 0.259 & 0.854 & \underline{4.54} & 0.039 & \underline{0.031} & 0.553 & 3.45 (3)\\
\textbf{SDC ($\alpha = 0.1$, $\lambda = 0.1$)} & \textbf{0.869} & \underline{1.82} & \textbf{0.039} & \textbf{0.044} & \textbf{0.173} & \underline{0.858} & \textbf{4.13} & \textbf{0.028} & 0.034 & 0.435 & 2.05 (1)\\
\bottomrule
\hline
\end{tabular}
}
\end{table*}
\footnotetext[1]{We follow the methodology of MICCAI'23~\cite{murugesan2023trust}, utilizing their publicly code~\cite{mur-github-url} and set NACL's kernel as "min" which is the lowest ECE from their experiments.}
\begin{table*}
\centering
\caption{\textbf{nnU-Net Segmentation and Calibration Results\protect\footnotemark[1].} 
Shown are discrimination metrics (DSC~$\uparrow$, HD~$\downarrow$) alongside calibration measures (ECE~$\downarrow$, CECE~$\downarrow$, pECE~$\downarrow$). Bold and underlined values highlight the best and second-best scores.}

\label{tab:quantitative_performance_nnunet}
\resizebox{\textwidth}{!}{%
\begin{tabular}{l|ccccc|ccccc|c}
\hline
\toprule
& \multicolumn{5}{c|}{\textbf{ACDC}} & \multicolumn{5}{c|}{\textbf{FLARE}} & {\textbf{Friedman}} \\
\hline 
\textbf{} & \textbf{DSC} & \textbf{HD} & \textbf{ECE} & \textbf{CECE} & \textbf{pECE} & \textbf{DSC} & \textbf{HD} & \textbf{ECE} & \textbf{CECE} & \textbf{pECE} & \textbf{Rank} \\
\hline
\midrule
DiceCE \cite{milletari2016v}  & 0.882 & 1.58 & 0.072 & 0.041 & 0.509 & 0.885 & 4.01 & 0.036 & 0.034 & 0.518 &  6.10 (7) \\
FL \cite{lin2017focal} ($\gamma = 3$)  & 0.872 & 1.60 & 0.089 & 0.065 & \underline{0.169} & 0.862 & 3.93 & 0.039 & 0.043 & 0.476 & 7.10 (9) \\
ECP \cite{pereyra2017regularizing} ($\lambda = 0.1$)  & 0.879 & 1.48 & 0.067 & 0.112 & 0.205 & 0.869 & 3.85 & 0.046 & 0.131 & 0.454 & 6.70 (8) \\
LS \cite{szegedy2016rethinking} ($\alpha = 0.1$)  & \textbf{0.885} & 1.46 & 0.062 & 0.057 & 0.170 & 0.866 & 4.25 & 0.059 & 0.051 & \textbf{0.316} & 5.30 (6) \\
$SVLS_{IPMI'21}$ \cite{islam2021spatially} ($\sigma = 2.0$) & 0.879 & 2.86 & 0.059 & 0.111 & 0.172 & 0.886 & \textbf{3.15} & 0.029 & 0.029 &  \underline{0.351} & 4.75 (5) \\
$MbLS_{CVPR'22}$ \cite{liu2022devil} ($m = 10$) & \underline{0.883} & 1.46 & 0.057 & 0.052 & 0.170 & 0.883 & 3.48 & 0.031 & 0.031 & 0.489 & 4.10 (2)  \\
$NACL_{MICCAI'23}$ \cite{murugesan2023trust} & 0.881 & 1.52 & 0.056 & 0.059 & 0.214 &  \underline{0.886} & 3.67 &  \underline{0.026} &  \underline{0.027} & 0.439 & 4.25 (4) \\
FCL \cite{liang2024calibrating} ($\gamma = 3$, $\lambda = 0.1$) & 0.882 & \textbf{1.26} & \textbf{0.032} & \textbf{0.035} & 0.179     & 0.879 & 3.89 & 0.036 & \textbf{0.026} & 0.525 & 4.10 (2)\\
\textbf{SDC ($\alpha = 0.1$, $\lambda = 0.1$)} & 0.880 & \underline{1.31} & \underline{0.037} & \underline{0.042} & \textbf{0.165} & \textbf{0.888} & \underline{3.46} & \textbf{0.023} & \textbf{0.026} & 0.416 & 2.30 (1) \\
\bottomrule
\hline
\end{tabular}
}
\end{table*}

\begin{table*}
    \centering
    \caption{Segmentation Performance Metrics Across Morphological Operations}
    \resizebox{0.9\textwidth}{!}{%
    \begin{tabular}{lcccccccc}
        \hline
        Metric & \multicolumn{1}{c}{\shortstack{Default \\ w/o}}& \multicolumn{1}{c}{\shortstack{Internal \\ Boundary}} & \multicolumn{1}{c}{\shortstack{External \\ Boundary}} & \multicolumn{1}{c}{Closing} & \multicolumn{1}{c}{Opening} & \multicolumn{1}{c}{Dilation} & \multicolumn{1}{c}{Erosion} & \multicolumn{1}{c}{\shortstack{Morphological \\ Gradient}} \\
        \hline
        DSC   &0.854& 0.863 & 0.861 & 0.861 & 0.849 & 0.857 & 0.862 & 0.861 \\
        HD    &2.932& 1.613 & 1.900 & 1.436 & 1.646 & 1.326 & 1.501 & 1.698 \\
        ECE   &0.068& 0.043 & 0.042 & 0.039 & 0.045 & 0.044 & 0.045 & 0.042 \\
        CECE  &0.061& 0.047 & 0.046 & 0.043 & 0.046 & 0.045 & 0.049 & 0.047 \\
        pECE  &0.281& 0.199 & 0.170 & 0.221 & 0.217 & 0.225 & 0.165 & 0.206 \\
        \hline
    \end{tabular}%
    }
    \label{tab:morphological_results}
\end{table*}
\begin{table*}
\centering
\caption{\textbf{U-Net Segmentation and Calibration Results on BraTS and Prostate\protect\footnotemark[1].} Bold and underlined values highlight the best and second-best scores.}
\label{tab:brats_prostate_8}
\resizebox{\textwidth}{!}{%
\begin{tabular}{l|ccccc|ccccc|c}
\toprule
& \multicolumn{5}{c|}{\textbf{BraTS}} 
& \multicolumn{5}{c|}{\textbf{Prostate}} 
& \textbf{Friedman} \\
\midrule
\textbf{Method} 
 & \textbf{DSC} & \textbf{HD} & \textbf{ECE} & \textbf{CECE} & \textbf{pECE}
 & \textbf{DSC} & \textbf{HD} & \textbf{ECE} & \textbf{CECE} & \textbf{pECE}
 & \textbf{Rank}\\
\midrule
DiceCE~\cite{milletari2016v}    
  & 0.739 & 18.73 & 0.246 & 0.159 & 0.556 
  & \underline{0.524} & 10.95 & 0.254 & 0.227 & 0.507  & 8.20 (9)\\
FL~\cite{lin2017focal} $(\gamma=3)$   
  & \underline{0.776} & 14.82 & 0.187 & 0.145 & 0.454
  & 0.520 & 9.93 & 0.225 & 0.205 & 0.424 
  & 4.40 (4)\\
ECP~\cite{pereyra2017regularizing} $(\lambda=0.1)$  
  & 0.742 & 16.55 & 0.232 & 0.151 & \underline{0.425} 
  & \textbf{0.526} & \underline{9.36} & 0.208 & \textbf{0.199} & \underline{0.311} 
  & 4.25 (3)\\
LS~\cite{szegedy2016rethinking} $(\alpha=0.1)$    
  & 0.752 & \textbf{12.14} & \underline{0.171} & 0.161 & \textbf{0.410 }
  & 0.501 & 10.88 & \underline{0.200} & 0.208 & 0.388  
  & 4.80 (6)\\
SVLS~\cite{islam2021spatially} $(\sigma=2.0)$ 
  & 0.761 & 14.62 & 0.176 & \textbf{0.140} & 0.466 
  & 0.481 & 10.29 & \textbf{0.186} & 0.216 & 0.326 
  & 4.15 (2)\\
MbLS~\cite{liu2022devil} $(m=10)$    
  & 0.746 & 16.02 & 0.218 & 0.148 & 0.452 
  & 0.508 & 10.34 & 0.223 & 0.218 & 0.381  
  & 6.05 (8)\\
NACL~\cite{murugesan2023trust}
  & 0.761 & 15.00 & 0.188& 0.144 & 0.451 
  & 0.480 & 10.29 & \textbf{0.186} & 0.216 & 0.326  
  & 4.55 (5)\\
FCL~\cite{liang2024calibrating}
  &0.754 & 16.68 & 0.214 &\underline{0.142} &0.447&0.506&10.74&0.251&0.221&0.335 &5.90 (7) \\
\textbf{SDC} $(\alpha=0.1, \lambda=0.1)$
  & \textbf{0.782}&\underline{13.45}&\textbf{0.169}&0.148&\underline{0.425}&0.516&\textbf{9.01}&0.230&\underline{0.202}&\textbf{0.305} & 2.70 (1) \\
\bottomrule
\end{tabular}
}
\end{table*}

\section{Results}
\label{sec:results_insights}

\noindent \textbf{Performance.} 
Tables~\ref{tab:quantitative_performance} and \ref{tab:quantitative_performance_nnunet} summarize segmentation and calibration outcomes for U-Net and nnU-Net, respectively. On \textbf{ACDC}, SDC attains the highest DSC (0.869) for U-Net and near-top DSC (0.880) for nnU-Net while maintaining competitive boundary delineation (HD). On \textbf{FLARE}, SDC achieves strong DSC (0.858 for U-Net and 0.888 for nnU-Net), outperforming most baselines. Notably, SDC registers the lowest ECE values across both datasets and backbones (e.g., 0.039 on ACDC with U-Net, 0.023 on FLARE with nnU-Net), indicating robust confidence alignment. Although SDC does not always yield the minimal pECE on FLARE (where LS occasionally excels), its overall Friedman rank remains the best for both U-Net and nnU-Net. This consistency underscores that precise anatomical segmentation and well-calibrated confidence estimates can be simultaneously achieved through our local calibration and SDF-based objectives. \textbf{BraTS} and \textbf{PROSTATE} results also prove it in Tab.~\ref{tab:brats_prostate_8}. We also observe that lower HD values are accompanied by lower calibration errors. An ablation study on morphological operation is provided in Tab.~\ref{tab:morphological_results}. 

\begin{figure}[t]
\centering
\includegraphics[width=\textwidth]{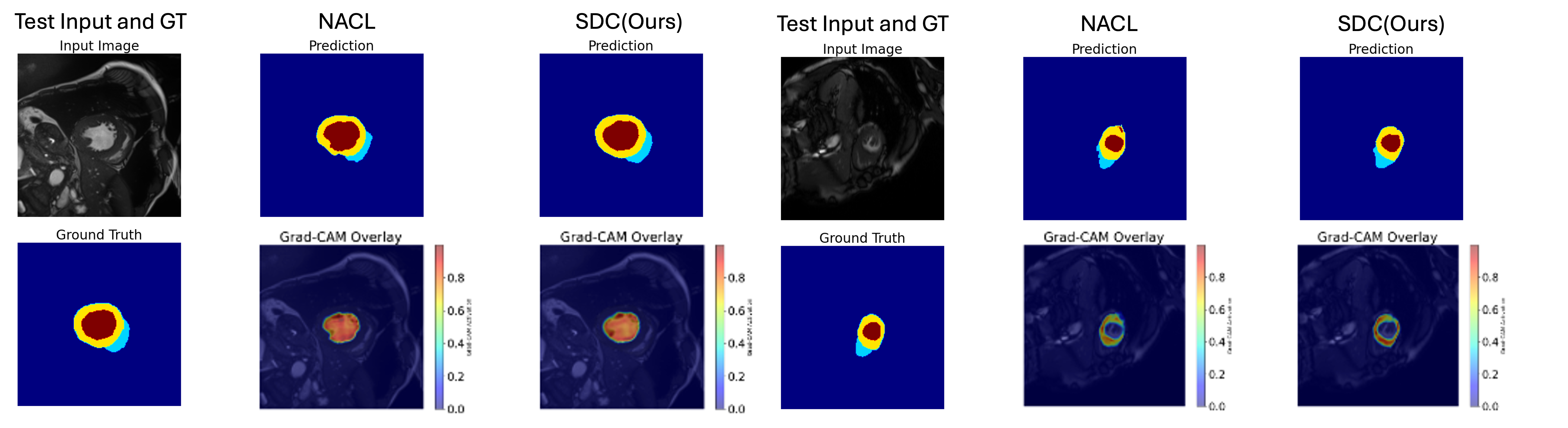}
\caption{Grad-CAM Heatmap, warm colors indicating regions the network relies upon most heavily for decisions. The predicted edge of SDC is more concise than NACL.} \label{grad_cam_visual}
\end{figure}

\noindent \textbf{Visualization.} 
Fig.~\ref{grad_cam_visual} illustrates representative slices predicted from the ACDC dataset by Grad-CAM~\cite{selvaraju2017grad}. It confirms accurate delineation of the left ventricle (red), myocardium (yellow), and right ventricle (cyan), with fewer boundary artifacts in SDC’s outputs. 

\section{Conclusion}
Medical image segmentation often demands high boundary accuracy, where small errors can have significant clinical ramifications. Discriminative losses (e.g., DiceCE) and even certain calibration-oriented methods (e.g., NACL) may yield strong results on select metrics but often lack consistency across all segmentation and calibration measures. By contrast, SDC—featuring local calibration and signed distance constraints—consistently improves both boundary delineation and confidence reliability. The visual evidence confirms that SDC better suppresses overconfident false positives at organ boundaries, providing probabilities that more faithfully mirror actual voxel-wise correctness.

\newpage
\bibliographystyle{splncs04}
\bibliography{mybibliography_miccai}

\begin{thebibliography}{10}
\providecommand{\url}[1]{\texttt{#1}}
\providecommand{\urlprefix}{URL }
\providecommand{\doi}[1]{https://doi.org/#1}

\bibitem{antonelli2022medical}
Antonelli, M., Reinke, A., Bakas, S., Farahani, K., Kopp-Schneider, A., Landman, B.A., Litjens, G., Menze, B., Ronneberger, O., Summers, R.M., et~al.: The medical segmentation decathlon. Nature communications  \textbf{13}(1), ~4128 (2022)

\bibitem{anwar2018medical}
Anwar, S.M., Majid, M., Qayyum, A., Awais, M., Alnowami, M., Khan, M.K.: Medical image analysis using convolutional neural networks: a review. Journal of medical systems  \textbf{42},  1--13 (2018)

\bibitem{Bakas2017AdvancingFeaturesJ}
Bakas, S., Akbari, H., Sotiras, A., Bilello, M., Rozycki, M., Kirby, J.S., Freymann, J.B., Farahani, K., Davatzikos, C.: Advancing the cancer genome atlas glioma {MRI} collections with expert segmentation labels and radiomic features. Scientific data  \textbf{4}(1),  1--13 (2017)

\bibitem{Bakas2018IdentifyingChallengeJ}
Bakas, S., Reyes, M., Jakab, A., Bauer, S., Rempfler, M., Crimi, A., Shinohara, R.T., Berger, C., Ha, S.M., Rozycki, M., et~al.: Identifying the best machine learning algorithms for brain tumor segmentation, progression assessment, and overall survival prediction in the brats challenge. arXiv preprint arXiv:1811.02629  (2018)

\bibitem{bernard2018deep}
Bernard, O., Lalande, A., Zotti, C., Cervenansky, F., Yang, X., Heng, P.A., Cetin, I., Lekadir, K., Camara, O., Ballester, M.A.G., et~al.: Deep learning techniques for automatic {MRI} cardiac multi-structures segmentation and diagnosis: is the problem solved? IEEE TMI  \textbf{37}(11),  2514--2525 (2018)

\bibitem{friedman1937use}
Friedman, M.: The use of ranks to avoid the assumption of normality implicit in the analysis of variance. Journal of the american statistical association  \textbf{32}(200),  675--701 (1937)

\bibitem{guo2017calibration}
Guo, C., Pleiss, G., Sun, Y., Weinberger, K.Q.: On calibration of modern neural networks. In: International conference on machine learning. pp. 1321--1330. PMLR (2017)

\bibitem{hricak2007imaging}
Hricak, H., Choyke, P.L., Eberhardt, S.C., Leibel, S.A., Scardino, P.T.: Imaging prostate cancer: a multidisciplinary perspective. Radiology  \textbf{243}(1),  28--53 (2007)

\bibitem{nnUNet}
Isensee, F., et~al.: nnu-net: a self-configuring method for deep learning-based biomedical image segmentation. Nature Methods  \textbf{18} (2020)

\bibitem{islam2021spatially}
Islam, M., Glocker, B.: Spatially varying label smoothing: Capturing uncertainty from expert annotations. In: Information Processing in Medical Imaging: 27th International Conference, IPMI 2021, Virtual Event, June 28--June 30, 2021, Proceedings 27. pp. 677--688. Springer (2021)

\bibitem{kingma2014adam}
Kingma, D.P., Ba, J.: Adam: A method for stochastic optimization. ICLR  (2015)

\bibitem{kumar2019verified}
Kumar, A., Liang, P.S., Ma, T.: Verified uncertainty calibration. Advances in Neural Information Processing Systems  \textbf{32} (2019)

\bibitem{lee2020structure}
Lee, H.J., Kim, J.U., Lee, S., Kim, H.G., Ro, Y.M.: Structure boundary preserving segmentation for medical image with ambiguous boundary. In: Proceedings of the IEEE/CVF conference on computer vision and pattern recognition. pp. 4817--4826 (2020)

\bibitem{liang2024calibrating}
Liang, W., Dong, C., Zheng, L., Li, Z., Zhang, W., Chen, W.: Calibrating deep neural network using euclidean distance. arXiv preprint arXiv:2410.18321  (2024)

\bibitem{lin2017focal}
Lin, T.Y., Goyal, P., Girshick, R., He, K., Doll{\'a}r, P.: Focal loss for dense object detection. In: Proceedings of the IEEE international conference on computer vision. pp. 2980--2988 (2017)

\bibitem{liu2022devil}
Liu, B., Ben~Ayed, I., Galdran, A., Dolz, J.: The devil is in the margin: Margin-based label smoothing for network calibration. In: Proceedings of the IEEE/CVF Conference on Computer Vision and Pattern Recognition. pp. 80--88 (2022)

\bibitem{Ma2021postrank}
Ma, X., Blaschko, M.B.: Meta-cal: Well-controlled post-hoc calibration by ranking. In: ICML (2021)

\bibitem{Menze2015TheBRATSJ}
Menze, B.H., et~al.: The multimodal brain tumor image segmentation benchmark (brats). IEEE Transactions on Medical Imaging  \textbf{34}(10),  1993--2024 (2015)

\bibitem{milletari2016v}
Milletari, F., Navab, N., Ahmadi, S.A.: V-net: Fully convolutional neural networks for volumetric medical image segmentation. In: 2016 fourth international conference on 3D vision (3DV). pp. 565--571. Ieee (2016)

\bibitem{mukhoti2020calibrating}
Mukhoti, J., Kulharia, V., Sanyal, A., Golodetz, S., Torr, P.H., Dokania, P.K.: Calibrating deep neural networks using focal loss. In: NeurIPS (2020)

\bibitem{murugesan2023trust}
Murugesan, B., Adiga~Vasudeva, S., Liu, B., Lombaert, H., Ben~Ayed, I., Dolz, J.: Trust your neighbours: Penalty-based constraints for model calibration. In: MICCAI. pp. 572--581 (2023)

\bibitem{mur-github-url}
Murugesan, B., Adiga~Vasudeva, S., Liu, B., Lombaert, H., Ben~Ayed, I., Dolz, J.: Trust your neighbours: Penalty-based constraints for model calibration. \url{https://github.com/Bala93/MarginLoss} (2023), accessed: 2025-01-13

\bibitem{murugesan2022calibrating}
Murugesan, B., Liu, B., Galdran, A., Ayed, I.B., Dolz, J.: Calibrating segmentation networks with margin-based label smoothing. Medical Image Analysis  \textbf{87},  102826 (2023)

\bibitem{murugesan2024neighbor}
Murugesan, B., Vasudeva, S.A., Liu, B., Lombaert, H., Ayed, I.B., Dolz, J.: Neighbor-aware calibration of segmentation networks with penalty-based constraints. arXiv preprint arXiv:2401.14487  (2024)

\bibitem{naeini2015ece}
Naeini, M.P., Cooper, G.F., Hauskrecht, M.: Obtaining well calibrated probabilities using bayesian binning. In: AAAI (2015)

\bibitem{pereyra2017regularizing}
Pereyra, G., Tucker, G., Chorowski, J., Kaiser, {\L}., Hinton, G.: Regularizing neural networks by penalizing confident output distributions. arXiv preprint arXiv:1701.06548  (2017)

\bibitem{pham2000current}
Pham, D.L., Xu, C., Prince, J.L.: Current methods in medical image segmentation. Annual review of biomedical engineering  \textbf{2}(1),  315--337 (2000)

\bibitem{rogowska2009overview}
Rogowska, J.: Overview and fundamentals of medical image segmentation. Handbook of medical image processing and analysis pp. 73--90 (2009)

\bibitem{ronneberger2015u}
Ronneberger, O., Fischer, P., Brox, T.: U-net: Convolutional networks for biomedical image segmentation. In: MICCAI. pp. 234--241 (2015)

\bibitem{sarvamangala2022convolutional}
Sarvamangala, D., Kulkarni, R.V.: Convolutional neural networks in medical image understanding: a survey. Evolutionary intelligence  \textbf{15}(1),  1--22 (2022)

\bibitem{selvaraju2017grad}
Selvaraju, R.R., Cogswell, M., Das, A., Vedantam, R., Parikh, D., Batra, D.: Grad-cam: Visual explanations from deep networks via gradient-based localization. In: Proceedings of the IEEE international conference on computer vision. pp. 618--626 (2017)

\bibitem{sox2024medical}
Sox, H.C., Higgins, M.C., Owens, D.K., Schmidler, G.S.: Medical decision making. John Wiley \& Sons (2024)

\bibitem{szegedy2016rethinking}
Szegedy, C., Vanhoucke, V., Ioffe, S., Shlens, J., Wojna, Z.: Rethinking the inception architecture for computer vision. In: CVPR (2016)

\bibitem{wang2022boundary}
Wang, R., Chen, S., Ji, C., Fan, J., Li, Y.: Boundary-aware context neural network for medical image segmentation. Medical image analysis  \textbf{78},  102395 (2022)

\bibitem{yeung2023calibrating}
Yeung, M., Rundo, L., Nan, Y., Sala, E., Sch{\"o}nlieb, C.B., Yang, G.: Calibrating the dice loss to handle neural network overconfidence for biomedical image segmentation. Journal of Digital Imaging  \textbf{36}(2),  739--752 (2023)

\bibitem{yeung2022unified}
Yeung, M., Sala, E., Sch{\"o}nlieb, C.B., Rundo, L.: Unified focal loss: Generalising dice and cross entropy-based losses to handle class imbalanced medical image segmentation. Computerized Medical Imaging and Graphics  \textbf{95},  102026 (2022)

\end{thebibliography}
%




\newpage
\appendix

\section{Appendix}
\subsection{Metric Computations}
\label{sec:appendix-metrics}

\subsubsection{Segmentation Performance}
We employ two widely-used segmentation metrics in medical imaging:

\paragraph{Dice Similarity Coefficient (DSC).}
For predicted segmentation \(\mathcal{P}\) and ground-truth segmentation \(\mathcal{G}\), DSC is given by
\begin{equation}
\label{eq:dsc}
\mathrm{DSC}(\mathcal{P}, \mathcal{G})
= \frac{2\,|\mathcal{P}\,\cap\,\mathcal{G}|}
       {|\mathcal{P}| + |\mathcal{G}|},
\end{equation}
where \(|\cdot|\) denotes the cardinality (i.e., number of pixels) in a set. A higher DSC indicates better overlap between prediction and ground truth.

\paragraph{95\% Hausdorff Distance (HD).}
We let \( d(p, \mathcal{G}) \) represent the Euclidean distance from a point \(p\) in \(\mathcal{P}\) to the closest point in \(\mathcal{G}\). Then the Hausdorff Distance is
\begin{equation}
\label{eq:hd}
\mathrm{HD}(\mathcal{P}, \mathcal{G})
= \max \Bigl\{
   \max_{p \,\in\, \mathcal{P}} \min_{g \,\in\, \mathcal{G}} d(p, g),\,
   \max_{g \,\in\, \mathcal{G}} \min_{p \,\in\, \mathcal{P}} d(g, p)
\Bigr\}.
\end{equation}
In practice, we report the 95\% HD, which discards extreme outliers by using the 95th percentile instead of the maximum.

\subsubsection{Calibration Performance}
To evaluate how well confidence estimates match actual correctness, we measure several calibration metrics:

\paragraph{Expected Calibration Error (ECE).}
Given foreground predictions binned into \(M\) intervals \(\{I_1, I_2, \dots, I_M\}\), let \(\bar{p}_m\) be the average confidence and \(\bar{a}_m\) the average accuracy in the \(m\)-th bin. The total number of foreground pixels is \(|\Omega|\). Then ECE is:
\begin{equation}
\label{eq:ece}
\mathrm{ECE}
= \sum_{m=1}^{M}
  \frac{|I_m|}{|\Omega|}
  \,\bigl|\bar{p}_m - \bar{a}_m\bigr|.
\end{equation}
\(\bar{p}_m\) and \(\bar{a}_m\) capture how over- or under-confident the model is within each bin.

\paragraph{Class-wise Expected Calibration Error (CECE).}
Let \(C\) be the total number of classes, and denote by \(\mathrm{ECE}_c\) the ECE for class \(c\). A simple way to aggregate per-class calibration is
\begin{equation}
\label{eq:cece}
\mathrm{CECE}
= \frac{1}{C} \sum_{c=1}^{C} \mathrm{ECE}_c,
\end{equation}
potentially discarding classes with negligible representation by applying a threshold on predicted probability (e.g., \(10^{-3}\)) as suggested in~\cite{kumar2019verified}.

\paragraph{Pixel-wise ECE (pECE).}
To capture more localized calibration errors, we group individual pixels by confidence rather than aggregating them globally. Suppose we use \(B\) bins \(\bigl[\beta_0, \beta_1), \dots, [\beta_{B-1}, \beta_B\bigr]\) over \([0,1]\). For each bin \(b\), let \(\bar{p}_b\) and \(\bar{a}_b\) be the mean predicted confidence and mean accuracy, respectively, and let \(\mathrm{FPConf}_b\) be the mean confidence of false positives. We define
\begin{equation}
\label{eq:pece_app}
\mathrm{pECE}
= \sum_{b=1}^{B}
  \frac{\bigl|(\bar{p}_b - \bar{a}_b) + \omega_{\mathrm{fp}} \,\mathrm{FPConf}_b\bigr|
  \,\cdot\,|\Omega_b|}{|\Omega|},
\end{equation}
where \(\Omega_b\) is the set of pixels with confidence in the \(b\)-th interval, \(|\Omega_b|\) is its cardinality, and \(|\Omega|\) is the total number of pixels. The term \(\omega_{\mathrm{fp}}\) penalizes excessive confidence in background regions, making pECE more sensitive to overconfident misclassifications.
\paragraph{Friedman Ranking}
To compare overall performance across multiple tasks or experimental configurations, we compute the Friedman ranking~\cite{friedman1937use}. Let \(\mathcal{M}\) be the set of metrics (e.g., DSC, HD, ECE, CECE, pECE.) measured on \(T\) separate models or methods. We assign each method a rank based on its relative score for each metric, then average ranks across all metrics in \(\mathcal{M}\). Formally, for each metric \(m\in \mathcal{M}\), we rank methods \(\{1,2,\ldots,T\}\) from best to worst; the Friedman statistic is then used to assess whether rank distributions differ significantly across methods.

\subsection{More Medical Datasets}
\begin{table*}[]
\centering
\caption{\textbf{Segmentation and Calibration Results (BraTS, Prostate)}. 
Each dataset has 5 metrics: DSC $\uparrow$, HD $\downarrow$, ECE $\downarrow$, CECE $\downarrow$, and pECE $\downarrow$.}
\label{tab:brats_prostate}
\resizebox{\textwidth}{!}{%
\begin{tabular}{l|ccccc|ccccc|c}
\toprule
& \multicolumn{5}{c|}{\textbf{BraTS}} 
& \multicolumn{5}{c|}{\textbf{Prostate}} 
& \textbf{Friedman} \\
\midrule
\textbf{Method} 
 & \textbf{DSC} & \textbf{HD} & \textbf{ECE} & \textbf{CECE} & \textbf{pECE}
 & \textbf{DSC} & \textbf{HD} & \textbf{ECE} & \textbf{CECE} & \textbf{pECE}
 & \textbf{Rank}\\
\midrule
DiceCE~\cite{milletari2016v}    
  & 0.739 & 18.73 & 0.246 & 0.159 & 0.556 
  & \underline{0.524} & 10.95 & 0.254 & 0.227 & 0.507  & 8.20 (9)\\
FL~\cite{lin2017focal} $(\gamma=3)$   
  & \underline{0.776} & 14.82 & 0.187 & 0.145 & 0.454
  & 0.520 & 9.93 & 0.225 & 0.205 & 0.424 
  & 4.40 (4)\\
ECP~\cite{pereyra2017regularizing} $(\lambda=0.1)$  
  & 0.742 & 16.55 & 0.232 & 0.151 & \underline{0.425} 
  & \textbf{0.526} & \underline{9.36} & 0.208 & \textbf{0.199} & \underline{0.311} 
  & 4.25 (3)\\
LS~\cite{szegedy2016rethinking} $(\alpha=0.1)$    
  & 0.752 & \textbf{12.14} & \underline{0.171} & 0.161 & \textbf{0.410 }
  & 0.501 & 10.88 & \underline{0.200} & 0.208 & 0.388  
  & 4.80 (6)\\
SVLS~\cite{islam2021spatially} $(\sigma=2.0)$ 
  & 0.761 & 14.62 & 0.176 & \textbf{0.140} & 0.466 
  & 0.481 & 10.29 & \textbf{0.186} & 0.216 & 0.326 
  & 4.15 (2)\\
MbLS~\cite{liu2022devil} $(m=10)$    
  & 0.746 & 16.02 & 0.218 & 0.148 & 0.452 
  & 0.508 & 10.34 & 0.223 & 0.218 & 0.381  
  & 6.05 (8)\\
NACL~\cite{murugesan2023trust}
  & 0.761 & 15.00 & 0.188& 0.144 & 0.451 
  & 0.480 & 10.29 & \textbf{0.186} & 0.216 & 0.326  
  & 4.55 (5)\\
FCL~\cite{liang2024calibrating}
  &0.754 & 16.68 & 0.214 &\underline{0.142} &0.447&0.506&10.74&0.251&0.221&0.335 &5.90 (7) \\
\textbf{SDC} $(\alpha=0.1, \lambda=0.1)$
  & \textbf{0.782}&\underline{13.45}&\textbf{0.169}&0.148&\underline{0.425}&0.516&\textbf{9.01}&0.230&\underline{0.202}&\textbf{0.305} & 2.70 (1) \\
\bottomrule
\end{tabular}
}
\end{table*}

\subsection{pECE Algorithm}
\label{pEce_Algo}
\begin{algorithm}[t]
\caption{Pixel-Wise Expected Calibration Error (pECE) with FP Offset}
\label{alg:pece}
\begin{algorithmic}[1]
\Require 
  \(\mathbf{p}\): predicted confidence map,
  \(\mathbf{y}\): ground-truth label map (same dimensions as \(\mathbf{p}\)),
  \(\texttt{bins}\): number of confidence intervals (default = 10),
  \(\texttt{fp\_weight}\): multiplier for penalizing false positives (default = 2.0)

\State \(\Delta \gets \texttt{linspace}(0, 1, \texttt{bins} + 1)\) 
  \Comment{Confidence bin edges}
\State \(\text{pECE} \gets 0.0\)
\State \(\text{total} \gets \texttt{numel}(\mathbf{p})\)
  \Comment{Total number of pixels}

\For{\(i \gets 1\) to \(\texttt{bins}\)}
    \State \(\Gamma_i \gets \Bigl\{(x,y) \,
      \Delta_{i-1} < \mathbf{p}[x,y] \leq \Delta_{i}\Bigr\}\)
      \Comment{All pixels whose confidence falls in bin \(i\)}
    \State \(\eta_i \gets |\Gamma_i|\)
    \If{\(\eta_i > 0\)}
        \State \(\widehat{p}_i \gets \frac{1}{\eta_i} \sum_{(x,y)\,\in\,\Gamma_i} \mathbf{p}[x,y]\)
          \Comment{Mean predicted confidence in bin \(i\)}
        \State \(\widehat{a}_i \gets \frac{1}{\eta_i} \sum_{(x,y)\,\in\,\Gamma_i} \mathbf{y}[x,y]\)
          \Comment{Mean accuracy in bin \(i\)}
        \State \(\Gamma_i^{\mathrm{fp}} \gets \{(x,y)\in \Gamma_i \mid \mathbf{y}[x,y] = 0\}\)
        \State \(\widehat{p}_i^{\mathrm{fp}} \gets 0.0\)
        \If{\(|\Gamma_i^{\mathrm{fp}}| > 0\)}
            \State \(\widehat{p}_i^{\mathrm{fp}} 
                \gets \frac{1}{|\Gamma_i^{\mathrm{fp}}|} 
                \sum_{(x,y)\,\in\,\Gamma_i^{\mathrm{fp}}} \mathbf{p}[x,y]\)
        \EndIf
        \State \(\text{offset} \gets \texttt{fp\_weight} \cdot \widehat{p}_i^{\mathrm{fp}}\)

        \State \(\text{pECE} \gets \text{pECE} \;+\; \frac{\eta_i}{\text{total}} \times 
            \Bigl|\bigl(\widehat{p}_i - \widehat{a}_i\bigr) + \text{offset}\Bigr|\)
    \EndIf
\EndFor

\State \textbf{return} \(\text{pECE}\)

\end{algorithmic}
\end{algorithm}

\section{Visualization}
\subsection{Grad-CAM Result}
\label{appendix:grad-cam}
\begin{figure}
\centering
\includegraphics[width=0.9\textwidth]{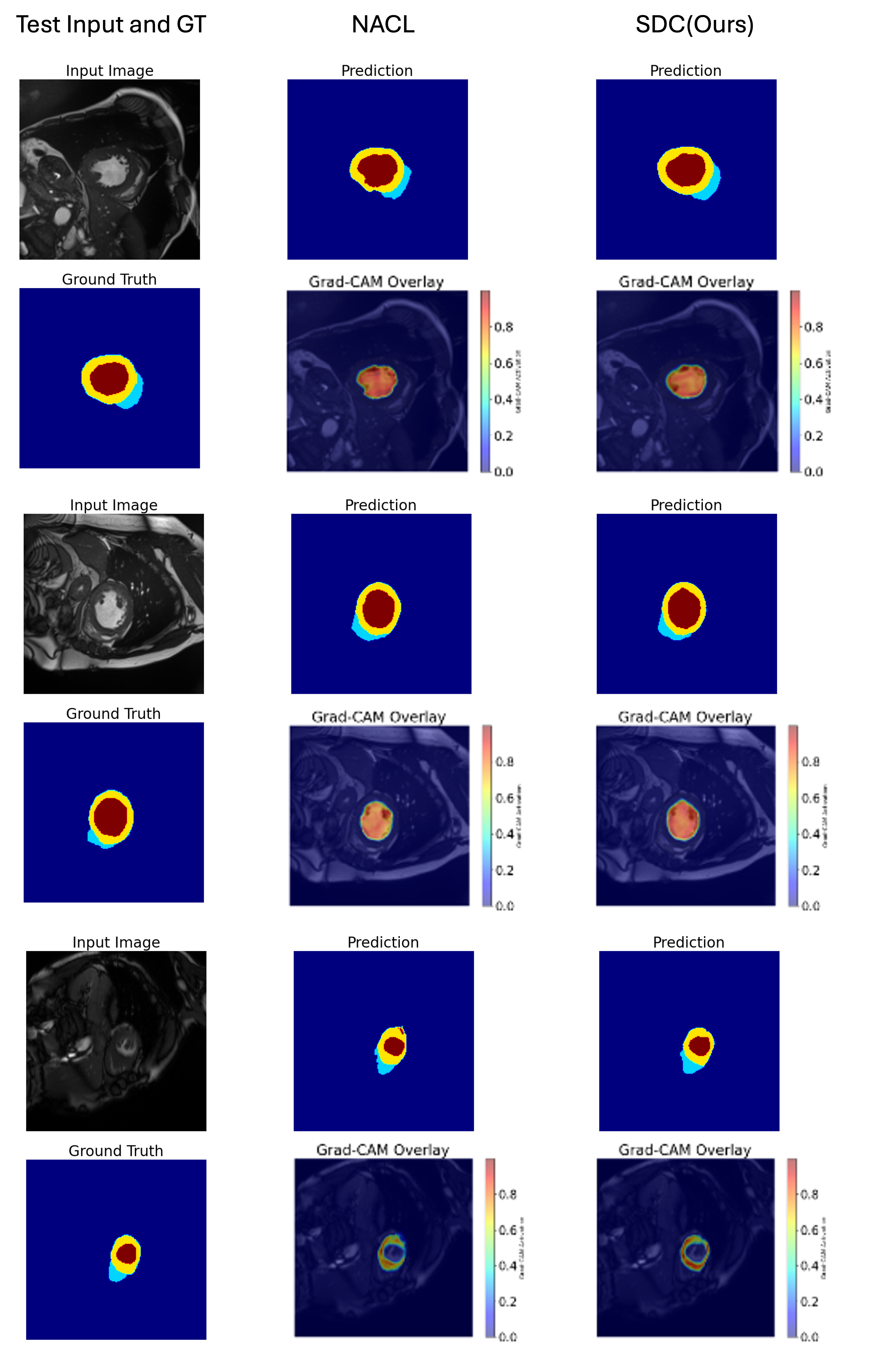}
\caption{Warmer colors (closer to red) in the Grad-CAM heatmap reflect higher feature importance, indicating regions the network relies upon most heavily for its segmentation decisions. Each structure (e.g., left ventricle, myocardium, right ventricle) 
is color-coded, facilitating a clear visual assessment of boundary quality and alignment with reference annotations.} \label{grad_cam}
\end{figure}

Fig.~\ref{grad_cam} is sample slices from the ACDC dataset illustrating our model's segmentation outputs and Grad-CAM activations. Each row displays (from left to right): the original MRI slice, the predicted segmentation map (second column), an alternative segmentation prediction for comparison (third column), the ground-truth annotations (fourth row within each block), and two Grad-CAM overlays. 

\subsection{Qualitative Insights}
\label{app:acdc_step_1627}
\begin{figure}
\includegraphics[width=\textwidth]{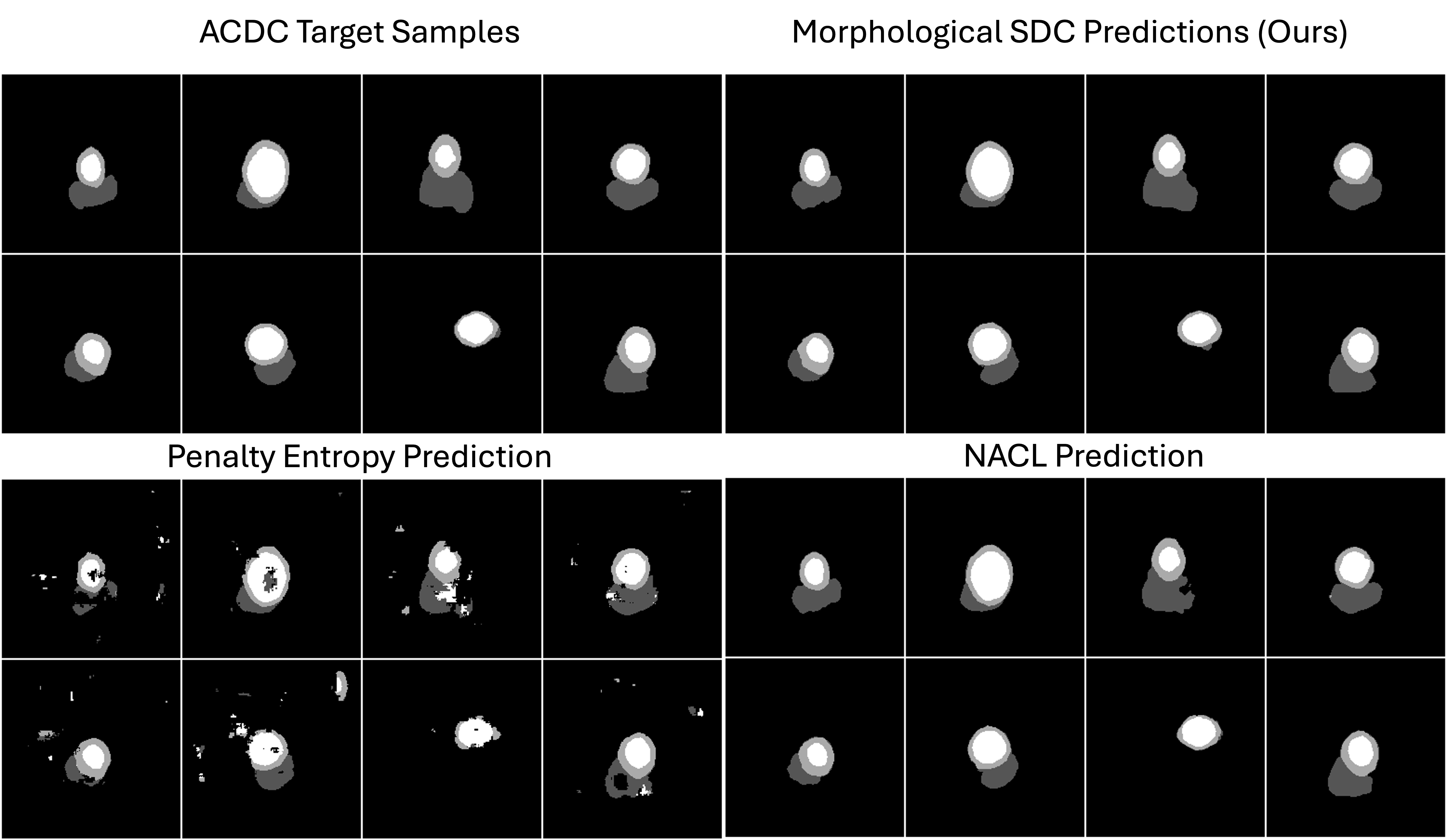}
\caption{\textbf{Qualitative Comparisons at Training Step 1627.}
Shown are ACDC target masks (first column) versus predictions from our Morphological NACL method (second column), Penalty Entropy (third column), and standard NACL (fourth column). Images are selected from the validation set at a mid-training checkpoint.} \label{fig:qualitative_acdc_comparison}
\end{figure}

Figure~\ref{fig:qualitative_acdc_comparison} compares segmentation results for four representative cardiac slices at training step 1627. The first column displays the ground-truth masks. The second column shows our Morphological NACL predictions, which generally exhibit precise boundaries and minimal false positives. By contrast, the Penalty Entropy method (third column) produces scattered artifacts, while standard NACL predictions (fourth column) often lack fine-grained boundary refinement. These qualitative observations align with the quantitative improvements we observe when incorporating morphological transforms.

\section{Formal Proof of SDF-Based Calibration Improvement}
\label{sec:sdf_calib_proof}

This appendix provides a formal proof that enforcing consistency between the predicted signed distance function (SDF) and the ground-truth SDF can theoretically improve calibration. The proof assumes an idealized logistic relationship between distance to the boundary and the probability of belonging to the object.

\subsection{Notation and Setup}

\begin{definition}[Ground-Truth SDF]
Let \(\Omega \subset \mathbb{R}^n\) be the image domain (\(n=2\) for 2D or \(n=3\) for 3D). Suppose we have a ground-truth segmentation \(\mathcal{O} \subset \Omega\) with boundary \(\partial \mathcal{O}\). The ground-truth SDF is defined as
\begin{equation}
  s^*(x) \;=\;
  \begin{cases}
    -\,d\bigl(x, \partial \mathcal{O}\bigr), & x \in \mathcal{O},\\[6pt]
    \;d\bigl(x, \partial \mathcal{O}\bigr), & x \notin \mathcal{O},
  \end{cases}
\end{equation}
where \(d(x, \partial \mathcal{O})\) denotes the Euclidean distance from \(x\) to the boundary \(\partial \mathcal{O}\).
\end{definition}

\begin{definition}[Predicted SDF and Probability]
A neural network outputs a predicted SDF \(s(x)\). We convert this to a segmentation probability 
\begin{equation}
  p(x) \;=\; \sigma\bigl(-\alpha\,s(x)\bigr),
\end{equation}
where \(\sigma(z) = \tfrac{1}{1+e^{-z}}\) is the logistic sigmoid, and \(\alpha > 0\) is a scaling factor. 
\end{definition}

\begin{definition}[Ideal (Intrinsic) Probability]
We define
\begin{equation}
  p^*(x) \;=\; \sigma\bigl(-\alpha\,s^*(x)\bigr).
\end{equation}
We say \(p^*(\cdot)\) is \emph{intrinsically calibrated} if, for any confidence value \(c\in [0,1]\), the set of points \(\{x : p^*(x)\approx c\}\) has a true label frequency close to \(c\). In other words, \(p^*\) accurately reflects the true likelihood of belonging to \(\mathcal{O}\).
\end{definition}

\begin{definition}[Calibration Error]
Let \(\mathrm{CalError}(p)\) denote a calibration metric such as ECE or pECE, measuring how well \(p\) aligns with empirical frequencies of belonging to \(\mathcal{O}\). A model is perfectly calibrated if \(\mathrm{CalError}(p) = 0\).
\end{definition}

\subsection{Lemma: Lipschitz Continuity of the Logistic Function}
\label{sec:logistic_lipschitz}

\begin{lemma}\label{lemma:logistic_lipschitz}
For all \(z_1, z_2 \in \mathbb{R}\), the logistic sigmoid \(\sigma(z) = \tfrac{1}{1+e^{-z}}\) satisfies
\begin{equation}
  \bigl|\sigma(z_1) - \sigma(z_2)\bigr| \;\le\; \tfrac{1}{4}\,\bigl|z_1 - z_2\bigr|.
\end{equation}
Consequently,
\begin{equation}
  \bigl|\sigma(-\alpha\,z_1) - \sigma(-\alpha\,z_2)\bigr|
  \;\le\; \tfrac{\alpha}{4}\,\bigl|z_1 - z_2\bigr|.
\end{equation}
\end{lemma}

\begin{proof}
The logistic function \(\sigma(z)\) has a global maximum slope of \(1/4\). By the mean value theorem, for \(z_1,z_2\in\mathbb{R}\),
\begin{equation}
  \bigl|\sigma(z_1) - \sigma(z_2)\bigr|
  \;\le\;
  \sup_{z \in [\min(z_1,z_2),\, \max(z_1,z_2)]} 
  \bigl|\sigma'(z)\bigr|
  \,\times\,
  |z_1 - z_2|
  \;\le\;
  \frac{1}{4}\,\bigl|z_1 - z_2\bigr|.
\end{equation}
For \(\sigma(-\alpha\,z)\), the slope is scaled by \(\alpha\), yielding the stated bound.
\end{proof}

\subsection{Bounding the Probability Error via SDF Error}

\begin{lemma}[SDF to Probability Discrepancy]\label{lemma:sdf_prob_discrepancy}
Suppose \(\|s - s^*\|_{\infty} \le \delta\). Then, for every \(x\in \Omega\),
\begin{equation}
  \bigl|p(x) - p^*(x)\bigr|
  \;\le\;
  \frac{\alpha\,\delta}{4},
\end{equation}
where \(p(x) = \sigma(-\alpha\,s(x))\) and \(p^*(x) = \sigma(-\alpha\,s^*(x))\).
\end{lemma}

\begin{proof}
By assumption, \(\max_{x\in\Omega}\bigl|s(x) - s^*(x)\bigr|\le\delta\). From Lemma~\ref{lemma:logistic_lipschitz},
\begin{equation}
  \bigl|\sigma\bigl(-\alpha\,s(x)\bigr)
         -\sigma\bigl(-\alpha\,s^*(x)\bigr)\bigr|
  \;\le\; 
  \frac{\alpha}{4}\,\bigl|s(x) - s^*(x)\bigr|
  \;\le\; 
  \frac{\alpha\,\delta}{4}.
\end{equation}
Thus \(\bigl|p(x) - p^*(x)\bigr|\le \tfrac{\alpha\,\delta}{4}\) for all \(x\in\Omega\).
\end{proof}

\subsection{Theorem: SDF Accuracy Improves Calibration}

\begin{theorem}\label{thm:sdf_cal_improve}
Let \(p^*\) be \emph{intrinsically calibrated}, i.e., \(\mathrm{CalError}(p^*) \approx 0\). Suppose \(\|s - s^*\|_{\infty}\le \delta\). Then the calibration error of \(p(x) = \sigma\bigl(-\alpha\,s(x)\bigr)\) satisfies
\begin{equation}
  \mathrm{CalError}\bigl(p(\cdot)\bigr)
  \;\le\;
  \mathrm{CalError}\bigl(p^*(\cdot)\bigr)
  \;+\;
  g(\delta),
\end{equation}
for some increasing function \(g(\delta)\) that satisfies \(g(\delta)\to 0\) as \(\delta\to 0\). If \(p^*\) is perfectly calibrated (\(\mathrm{CalError}(p^*)=0\)) and \(s\to s^*\) uniformly, then \(\mathrm{CalError}(p)\to 0\) as well.
\end{theorem}

\begin{proof}
\quad

\noindent \textbf{(1) Bounding the Probability Difference.}  
By Lemma~\ref{lemma:sdf_prob_discrepancy}, if \(\|s - s^*\|_{\infty}\le\delta\), then \(\|p - p^*\|_{\infty}\le (\alpha\,\delta)/4\). Hence each pixel’s predicted probability differs from the ideal probability by at most \(\frac{\alpha\,\delta}{4}\).

\noindent \textbf{(2) Effect on Bin-Based Calibration Metrics.}  
Calibration error (ECE or pECE) typically involves partitioning predicted confidences into bins \(\{b_1,b_2,\dots\} \\ \subset [0,1]\). If \(\bigl|p(x)-p^*(x)\bigr|\le \tfrac{\alpha\,\delta}{4}\), a pixel can move from one bin to another only if the bin boundaries lie within \(\tfrac{\alpha\,\delta}{4}\) of its ideal confidence. Thus the proportion of points in each bin shifts by an amount bounded by \(\tfrac{\alpha\,\delta}{4}\).

\noindent \textbf{(3) Decomposition of the Error.}  
If \(p^*\) is intrinsically calibrated, then each bin’s empirical frequency of truly positive pixels is very close to the bin’s midpoint. Under \(p\), the difference in bin membership (and thus the difference in empirical frequencies) depends on how many points transition into or out of each bin due to the shift \(\|p-p^*\|_{\infty}\). This shift is in turn bounded by \(\tfrac{\alpha\,\delta}{4}\), so the net effect on the calibration metric is captured by an increasing function \(g(\delta)\) with \(g(0)=0\).

\noindent \textbf{(4) Conclusion.}  
Hence,
\begin{equation}
  \mathrm{CalError}\bigl(p(\cdot)\bigr)
  \;\le\;
  \mathrm{CalError}\bigl(p^*(\cdot)\bigr)
  \;+\;
  g(\delta).
\end{equation}
As \(\delta\to 0\), \(\mathrm{CalError}(p)\to \mathrm{CalError}(p^*)\). If \(p^*\) is perfectly calibrated, then \(\mathrm{CalError}(p^*)=0\) implies \(\mathrm{CalError}(p)\to 0\).
\end{proof}

\subsection{Discussion and Practical Relevance}

The SDF loss term encourages the network to learn a predicted SDF \(s(x)\) that closely approximates the true SDF \(s^*(x)\). By mapping \(s(x)\) through a smooth logistic function, we avoid artificially “saturating” the confidence near 0 or 1 in boundary regions where genuine uncertainty is high.

Medical images frequently exhibit partially defined or fuzzy boundaries. Standard losses may over-penalize ambiguous pixels, driving probabilities to extremes. Enforcing SDF fidelity better preserves intermediate probabilities, aligning predictions with the actual likelihood of belonging to the target region. However, this result depends on the assumption that \(p^*(x)=\sigma(-\alpha\,s^*(x))\) is perfectly (or nearly) calibrated. Real-world data may introduce noise, label ambiguities, and artifacts that deviate from this ideal. Nonetheless, Thm.~\ref{thm:sdf_cal_improve} provides a theoretical foundation justifying the use of SDF for improved calibration.

By bounding the error \(\|s - s^*\|\), we ensure that \(p(x)\) remains close to \(p^*(x)\). Under an intrinsically calibrated ground-truth logistic mapping, this yields a provable reduction in calibration error. Therefore, introducing an SDF term in the loss function fosters more reliable probability estimates, mitigating overconfidence near boundaries and improving ECE/pECE metrics.

\section{Faster Convergence}
\label{app:faster_conv}
\begin{figure}
\includegraphics[width=\textwidth]{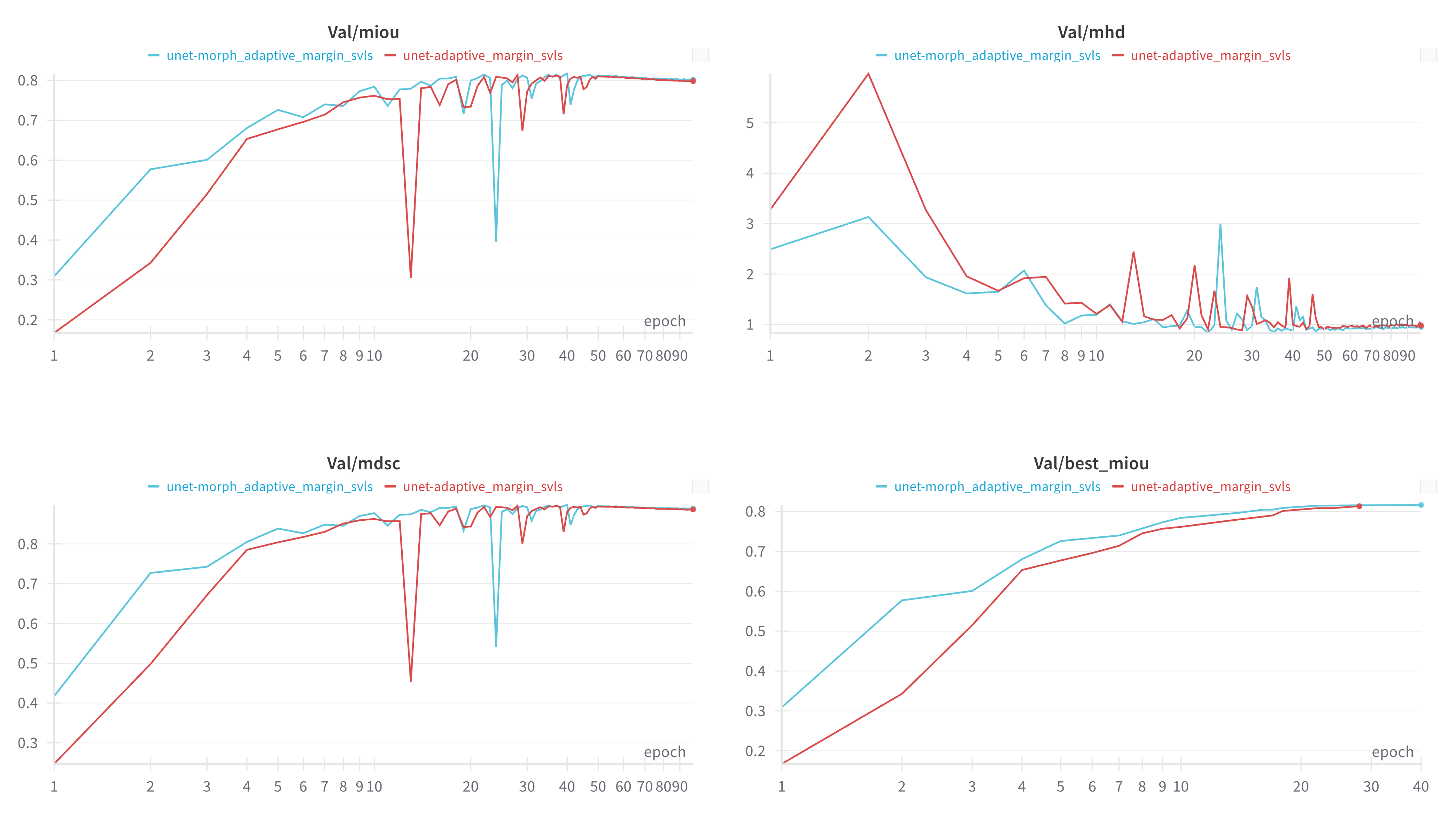}
\caption{\textbf{Faster Convergence with Morphological Transforms.} 
    Shown are validation curves (Scaled) for mean IoU, mean HD, mean DSC, and best mean IoU by U-Net. } \label{val_4m}
\end{figure}

In practice, morphological transforms help filter out label noise and refine boundary regions, creating more stable training signals early on. As illustrated in Figure~\ref{val_4m}, the baseline NACL (red) takes more epochs to stabilize and often displays oscillations in mean IoU and DSC. By contrast, our approach applied morphological transforms to NACL (blue) converges faster to higher values on all metrics—mean IoU, mean DSC, and even lower mean HD—demonstrating the advantage of cleaner boundary supervision. Our method SDC achieves the best IoU validation faster from other losses as well.

\section{Batch Runtime and Complexity}
\begin{figure}
\includegraphics[width=\textwidth]{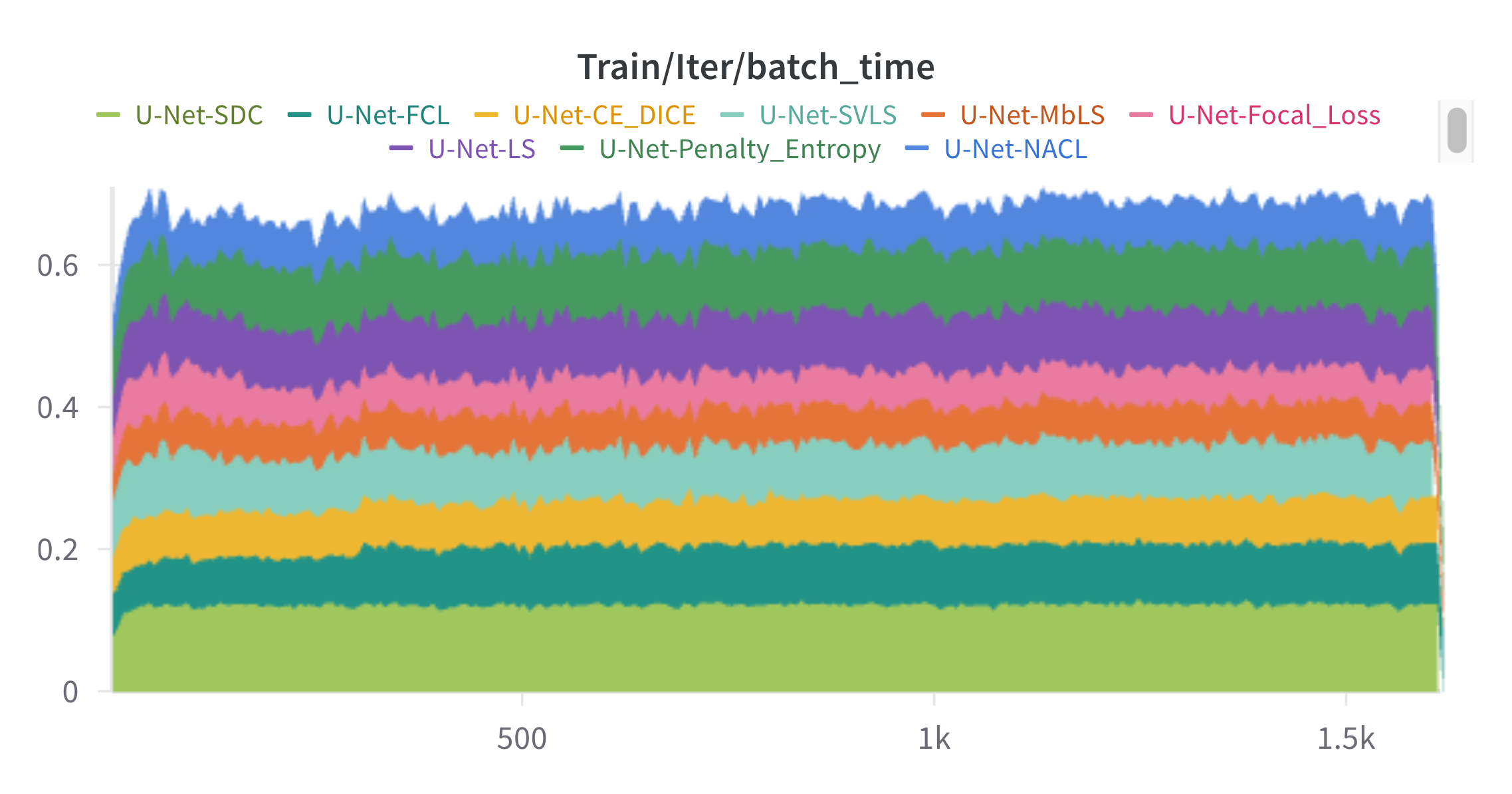}
\caption{Iteration of batch time per step by U-Net. } \label{time_comp}
\end{figure}

Each colored band represents the average time per training iteration (batch) for one of the listed U-Net variants, differing only by their loss functions in Fig.~\ref{time_comp}. 

\section{Ablation Study on Morphological Operation}
\label{app:ablation}
This experiment evaluates various segmentation quality metrics across different morphological operations applied to the segmented outputs. Default setting is NACL method without applying any operations.
\begin{table}[h]
    \centering
    \caption{Segmentation Performance Metrics Across Morphological Operations}
    \resizebox{\textwidth}{!}{%
    \begin{tabular}{lcccccccc}
        \hline
        Metric & \multicolumn{1}{c}{\shortstack{Default \\ w/o}}& \multicolumn{1}{c}{\shortstack{Internal \\ Boundary}} & \multicolumn{1}{c}{\shortstack{External \\ Boundary}} & \multicolumn{1}{c}{Closing} & \multicolumn{1}{c}{Opening} & \multicolumn{1}{c}{Dilation} & \multicolumn{1}{c}{Erosion} & \multicolumn{1}{c}{\shortstack{Morphological \\ Gradient}} \\
        \hline
        DSC   &0.854& 0.863 & 0.861 & 0.861 & 0.849 & 0.857 & 0.862 & 0.861 \\
        HD    &2.932& 1.613 & 1.900 & 1.436 & 1.646 & 1.326 & 1.501 & 1.698 \\
        ECE   &0.068& 0.043 & 0.042 & 0.039 & 0.045 & 0.044 & 0.045 & 0.042 \\
        CECE  &0.061& 0.047 & 0.046 & 0.043 & 0.046 & 0.045 & 0.049 & 0.047 \\
        pECE  &0.281& 0.199 & 0.170 & 0.221 & 0.217 & 0.225 & 0.165 & 0.206 \\
        \hline
    \end{tabular}%
    }
    \label{tab:morphological_results}
\end{table}

\section{Trade-off}
\begin{figure}
\includegraphics[width=\textwidth]{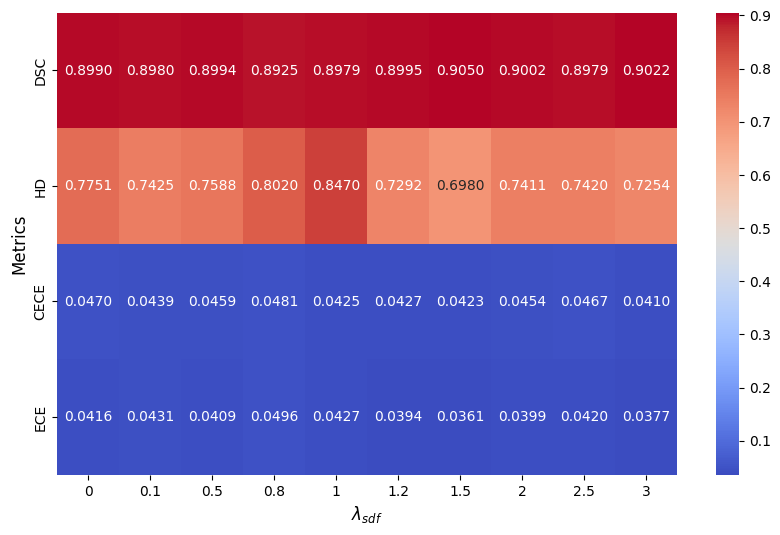}
\caption{Varied $\lambda_{sdf}$ for SDC by U-Net (ACDC). } \label{lambda_metrcis}
\end{figure}

The heatmap (Fig.~\ref{lambda_metrcis}) provides a clear representation of how MDSC, MHD, CECE, and ECE change with different values of $\lambda_{sdf}$. As $\lambda_{sdf}$ increases, DSC generally remains stable with slight fluctuations, peaking around 1.5 and 3.0. HD shows a notable increase at $\lambda_{sdf}$ = 1, indicating worse boundary performance, but improves at higher and lower values. Both CECE and ECE, which measure calibration error, exhibit variability, with ECE peaking at 0.8, suggesting that improper lambda selection may negatively impact calibration. Overall, the results suggest that choosing an optimal $\lambda_{sdf}$ (around 1.5 or 3.0) may help balance segmentation performance and calibration quality for ACDC.

\end{document}